\documentclass[prl, nofootinbib, twocolumn, preprintnumbers,amsmath,amssymb, superscriptaddress]{revtex4}

\usepackage{mathrsfs}%hanamoji
\usepackage{graphicx}% Include figure files
\usepackage{dcolumn}% Align table columns on decimal point
\usepackage{bm}% bold math
\usepackage{color}

\newcommand{\be}{\begin{equation}}
\newcommand{\ee}{\end{equation}}

\newcommand{\Ps}{P_s(n)}

\newcommand{\dng}{\dot n_g}

\begin{document}

\title{Experimental realization of a Szilard engine with a single electron}

\author{J. V. Koski \footnote{email jonne.koski@aalto.fi}}
\affiliation{Low Temperature Laboratory (OVLL), Aalto University, POB 13500, FI-00076 AALTO, Finland}
\author{V. F. Maisi}
\affiliation{Low Temperature Laboratory (OVLL), Aalto University, POB 13500, FI-00076 AALTO, Finland}
\affiliation{Centre for Metrology and Accreditation (MIKES), P.O. Box 9, 02151 Espoo, Finland}
\author{J. P. Pekola \footnote{email jukka.pekola@aalto.fi}}
\affiliation{Low Temperature Laboratory (OVLL), Aalto University, POB 13500, FI-00076 AALTO, Finland}
\author{D. V. Averin}
\affiliation{Department of Physics and Astronomy, Stony Brook University, SUNY, Stony Brook, NY 11794-3800, USA}

\date{\today}

\begin{abstract}

The most succinct manifestation of the second law of thermodynamics is the limitation the Landauer principle imposes on the amount of heat a Maxwell demon (MD) can convert into free energy per single bit of information obtained in a measurement. We suggest and experimentally realize a reversible electronic MD based on a single-electron box operated as a Szilard engine, providing the first demonstration of this limitation: extraction of $k_BT\ln2$ of heat from the reservoir at temperature $T$ per one bit of created information. The information is encoded in the position of an extra electron in the box.

\end{abstract}

%\keywords{ keywords}%Use showkeys class option if keyword
                              %display desired
\maketitle

The work of Maxwell suggesting what is now known as the ``Maxwell demon'' (MD)~\cite{MD}, which was quantified later on by Szilard~\cite{Szilard-29} initiated interest into the relationship between information and thermodynamics, see e.g., \cite{Brillouin, Bennett,Sagawa2008,Sagawa2010}. MD extracts heat from a thermal reservoir at temperature $T$ by observing a thermodynamic system to make a spontaneous, thermally-induced, transition into a state with larger-than-average free energy (either because of a larger internal energy or a smaller entropy) and using the feedback to collect this extra free energy as work. Szilard demonstrated that by obtaining a single bit of information as a measurement result of the state of the system, one could collect up to $k_B T \ln 2$ useful work, where  $k_B$ is the Boltzmann constant. Such a direct conversion of heat into work would by itself violate the second law of thermodynamics, because both the measurement and the feedback part of MD operation can in principle be done reversibly, without generating any extra entropy. In particular, classical reversible measurement can be viewed as a process of copying the state of the system into the memory of the detector. This means that the only fundamentally unavoidable thermodynamic costs of conversion of heat into work by a reversible MD is the creation of information about the state of the measured system. According to the Landauer principle \cite{Landauer61,Landauer88,Berut12}, erasure of this information generates at least the extracted amount of heat, $k_BT\ln2$ per bit, restoring the agreement with the second law.

\begin{figure}[h!t]
	\centering
	\includegraphics[width=0.797\columnwidth]{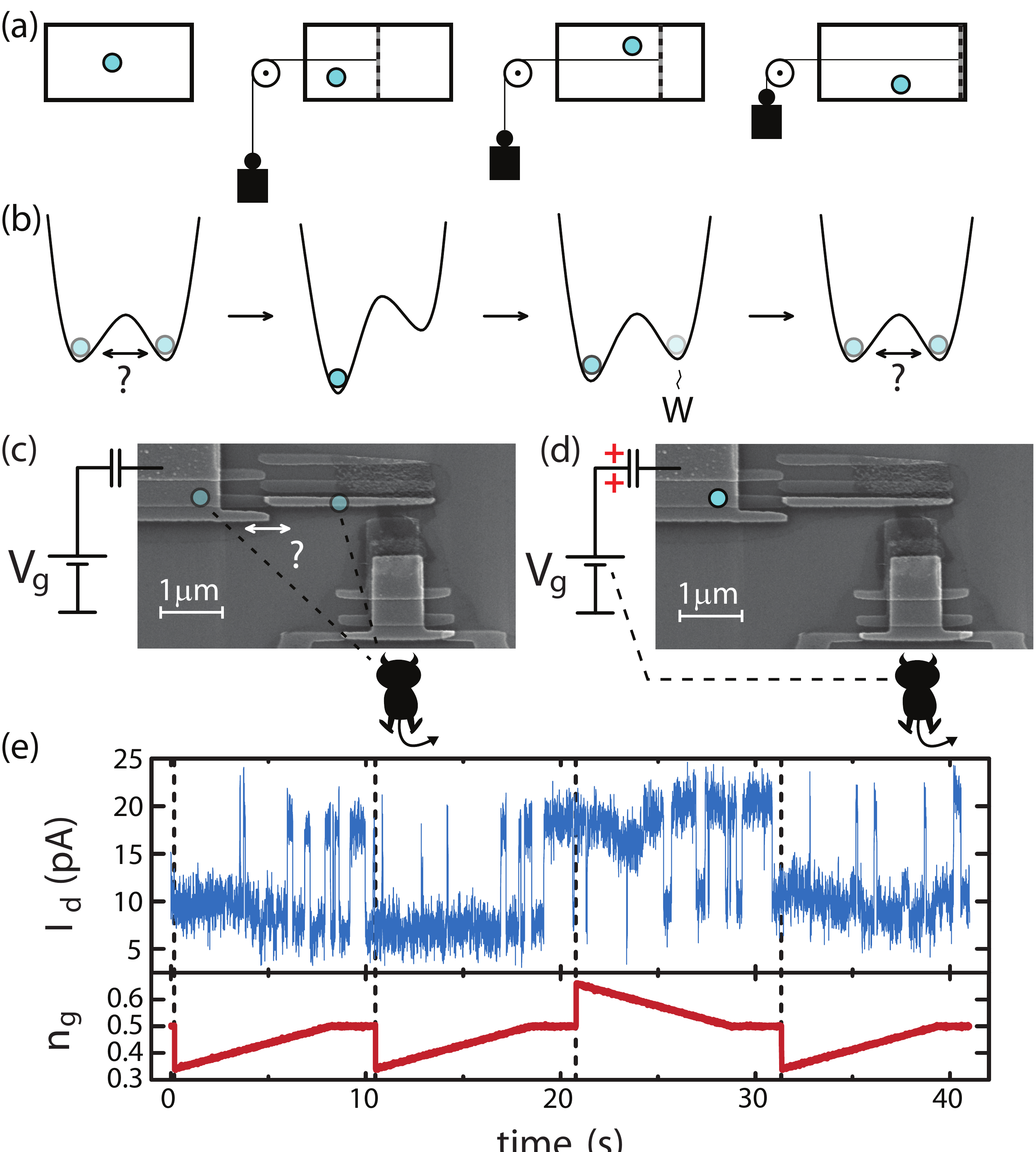}
	\caption{\label{fig:device} Szilard Engine.
(a) The original proposal of "Szilard engine": a box containing a single molecule is split into two equal sections (top left).
The section holding the molecule is allowed to expand up to the full volume of the box (top right). Then the partition is introduced again and the process repeats.
(b) Sketch of the energy diagrams allowing a similar cycle in the single-electron box (SEB). Work is extracted when the particle is thermally excited to the higher energy state.
(c) Experimental realization of the Szilard engine as SEB. An excess electron is located on one of the two metallic islands, corresponding to the first step on the panels (a) and (b).
(d) The measurement and feedback parts of our Maxwell demon operation. An SET electrometer on bottom detects the electron, while the gate voltage $V_g$ is applied to control the tunneling of the extra electron (to ``move the wall'') trapping it capacitively. As $V_g$ is slowly driven back to the original setup in (c), the net extracted work $k_B T \ln 2$ is produced by thermal activations as indicated in third step of panel (b). (e) A time trace of the excess electron location, signaled by the SET current $I_d$. The bottom trace shows the applied gate-voltage  signal that provides feedback. Here $n_g = C_g V_g/e$ with $C_g$ being  the coupling capacitance between the gate electrode and the gated box island.}
\end{figure}

While these general principles of MD operation are well understood in theory (see, e.g., the recent discussions \cite{Horowitz11,Mandal12,Barato13}), only few experimental realizations of a MD exist \cite{Toyabe10}, and thus far none demonstrates a quantitative connection between the MD output and the obtained information. The goal of this work is to suggest and realize a system that demonstrates explicitly the extraction of $k_BT\ln2$ of heat from a thermal reservoir by a MD per one bit of created information. The operating cycle we employ is close to the thought experiment suggested by Szilard, which illustrated the MD operation using as the working system measured and feedback-controlled molecule in a box. Panels in Fig.~\ref{fig:device}A, from left to right, show the steps of the operation of such a Szilard engine. The molecule is in equilibrium at temperature $T$, and the box is divided initially into two equal sections. After the measurement establishes which section the molecule is in, it is allowed to expand into the full volume lifting a weight tied to the dividing wall, thus extracting work from the thermal molecule. Then a dividing wall is introduced again and the cycle repeats. At the beginning of each cycle, the molecule has equal probabilities to be on the right or on the left, so that the measurement produces precisely one bit of information per cycle. As a result, in the reversible limit, the average extracted work per cycle reaches the fundamental maximum of $k_B T \ln 2$.

Our experimental realization of this cycle is shown in Fig.~\ref{fig:device}C. Its main element is the single electron box (SEB)~\cite{Averin86,Buttiker87,Lafarge91} which consists of two small metallic islands connected by a tunnel junction. The SEB is maintained at the dilution-refrigerator temperatures in the 0.1 K range. Physically, there are two main differences between the SEB and the original single-molecule Szilard engine. The electrodes of the box contain electron gas of a large number of electrons, and not just one particle. Consequently, what is being manipulated in the engine operation is not this single particle but the charge configuration of the box, which is determined by the position of one extra electron. Also, this manipulation is achieved not by partitioning and reconnecting the electrodes (which for the SEB would correspond to the modulation of the conductance of the tunnel junction connecting the islands) but by changing the potential difference between the electron gases in the two islands. Apart from these differences, the engine
follows the steps (illustrated with the potential profiles in Fig.~\ref{fig:device}B) similar to operation of the original Szilard engine. Potential difference between the islands is controlled by the gate voltage $V_g$ applied to one of them. Initially, $V_g$ is such that the extra electron is found equally likely on either of the islands (Fig.~\ref{fig:device}C). This ``degeneracy point'' is realized when the gate-offset charge $n_g= C_g V_g/e$, where $C_g$ is the capacitance between the gate and the SEB, is half integer. A single electron transistor (SET) electrometer, which can be seen on the bottom right in Figs.~\ref{fig:device}C and~\ref{fig:device}D, detects which island the electron is on. Then, $n_g$ is changed rapidly to capture electron on the corresponding island by increasing the energy required for tunneling out. Finally, $n_g$ is moved slowly back to the initial degeneracy value, extracting energy from the heat bath in the process, and completing the cycle. An example of four such consecutive experimental cycles is shown in Fig.~\ref{fig:device}E. Dotted vertical lines denote the time when the measurement is performed. We observe that the feedback signal indeed locks the extra electron to the measured state (parts of the trace in the upper panel in Fig.~\ref{fig:device}E with no jumps), but the charge starts to hop again when $n_g$ is moved towards the degeneracy point.

More quantitatively, the working space of the engine is spanned by the number $n$ of excess electrons on one of the box islands, while equilibrium electron gas in the box islands plays the role of the thermal reservoir at temperature $T$. Since there is only capacitive coupling between the box and the rest of the circuitry, electron tunneling takes place only between the two box islands. Therefore, the total electric charge on the two islands is conserved, and the state with $n$ excess electrons on one island has $-n$ excess electrons on the other island, as in a regular capacitor made of two electrodes. The internal energy of the engine is given then by the charging energy of these states, $E_n = E_c(n-n_g)^2$, averaged over their occupation probabilities $p_n$. Here $E_c = e^2/2 C_{tot}$ is the usual charging energy of the total capacitance $C_{tot}$ between the box islands. In the low-temperature regime relevant for this work, the charge dynamics is reduced to the two states, $n = 0,1$. Thermodynamics of the engine cycle described above qualitatively is characterized quantitatively~\cite{Averin11} by (i) the work done by the gate voltage source, $W = -\int \frac{d E_n(n_g)}{dn_g} dn_g$, and (ii) the heat $Q$ transferred to the electron gas of the box islands, i.e. to the thermal reservoir, by electron tunneling events. Note that electron tunneling events which change the charge state $n$ make the integral in the expression for work $W$ dependent on the specific realization of the history of the tunneling transitions. Each tunneling event produces the heat $Q = \pm(E_0(n_g) - E_1(n_g)) = \pm E_c (2 n_g - 1)$, where the plus sign describes the $n:0\to 1$ transitions, the minus sign - $n:1\to 0$ transitions. These relations enable us to measure directly the heat $Q$ transferred to the reservoir, as was done previously in \cite{Saira12,Koski13}, by detecting the electron tunneling events in real time and evaluating the corresponding energy difference, $E_0 - E_1$, at the moments of these events.

In the closed cycle of our experiment, energy conservation makes the total heat $-Q$ extracted from the reservoir equal to the work $-W$ extracted from the engine. The cycle starts with the SEB at degeneracy, and at this point, the charge state is measured by the external SET detector. One bit of information represented by the (equally probable) position of the extra electron on one or the other island of the box is copied into the detector and stored for the subsequent feedback process, where it is used to determine the polarity of the rapid gate-voltage drive. If the box is found in the state $n=0$, the gate voltage is changed rapidly so that the offset charge $n_g$ changes from the degeneracy value $n_g=1/2$ to $n_g=0$, if the measured state is $n=1$, $n_g$ changes from $n_g=1/2$ to $n_g=1$. Such a rapid feedback drive traps the electron to the measured state. Ideally, this drive is so fast that no electron transitions have a chance to occur during it and, as a result, no heat is transferred to the reservoir. 
%It should also be mentioned that for technical reasons, the SET detector in our  experiments is constantly active, so that we can keep track of specific realization of the electron tunneling pattern $n(t)$ and determine $Q$ and $W$ in the way discussed above. However, only a single bit of information about the initial charge state is used directly in the operation of our device.

The final part of the engine cycle is the quasi-static reversible ramp which returns the box to the degeneracy. Reversible nature of this ramp, combined with the absence of heat dissipation in the rapid feedback drive discussed in the previous paragraph, make the total operation cycle of our MD ideally reversible. Such reversibility distinguishes SEB setup in this work from other proposed electronic MDs \cite{Averin11b,Strasberg13,Bergli13} and is important for establishing the link between the extracted heat and information. Explicitly, the heat $Q$
extracted from the reservoir in the quasi-static ramp can be found
by considering the change of the total entropy $S$ of the box. This change, $\Delta S = \Delta S_r+\Delta S_{ch}$, consists of the standard entropy change of the thermal reservoir in equilibrium at temperature $T$ due to heat flow into it, $\Delta S_r = Q/T$, and the change of the Boltzmann entropy of the charge states
\be S_{ch} =-k_B \sum_n p_n \ln p_n \, . \label{eq:S} \ee
Using the standard rate equation for the evolution of the occupation probabilities $p_n$ \cite{Averin86}, one can find the rate of change of entropy $S$ due to electron tunneling in a general evolution process as
\be \frac{\partial S}{\partial t}
= \frac{1}{2} \sum_{n,m} \ln \Big[\frac{p_n \Gamma_{mn} }{p_m \Gamma_{nm}}\Big](p_n \Gamma_{mn} - p_m \Gamma_{nm} ) \, ,
\label{entropy} \ee
where $\Gamma_{mn}$ is the rate of electron tunneling from state $n$ to $m$. The tunneling rates satisfy the detailed-balance condition,
$\Gamma_{mn} =\Gamma_{nm}\exp\{(E_n-E_m)/k_BT\}$ (see Supplementary material for details). Equation (\ref{entropy}) shows that $S$ never decreases, and remains constant in the fully adiabatic evolution, when the probabilities $p_n$ maintain local equilibrium, $p_n\propto \exp\{-E_n(t)/k_BT\}$ and the detailed balance condition ensures that the probability fluxes vanish: $p_n \Gamma_{mn} =p_m \Gamma_{nm}$. In this case, the total entropy is conserved, $\Delta S=0$, and the two components of $S$ change in the opposite directions $\Delta S_r=Q/T=-\Delta S_{ch}$. Thus, the heat $Q$ extracted from the reservoir is determined by the change of the entropy of the charge states,
\be
Q= -T \Delta S_{ch}.
\label{heat} \ee
For the perfectly quasistatic ramp in our Szilard engine cycle, which brings the box from the definite charge state to the degeneracy point, this gives $Q=-k_B T \ln 2$. Qualitatively, this means that we are extracting $k_B T \ln 2$ of heat from the reservoir by creating a bit of information determined by the electron position on one or the other island of the SEB. In terms of work $W$, it is first extracted from the box by rapid lowering of the potential, as sketched in Fig.~\ref{fig:device} B. Work is then applied to drive the box back to the  degeneracy, however the required work is lowered by the amount of heat $k_B T \ln 2$ absorbed from the thermal bath.

\begin{figure}
\includegraphics[width=\columnwidth]{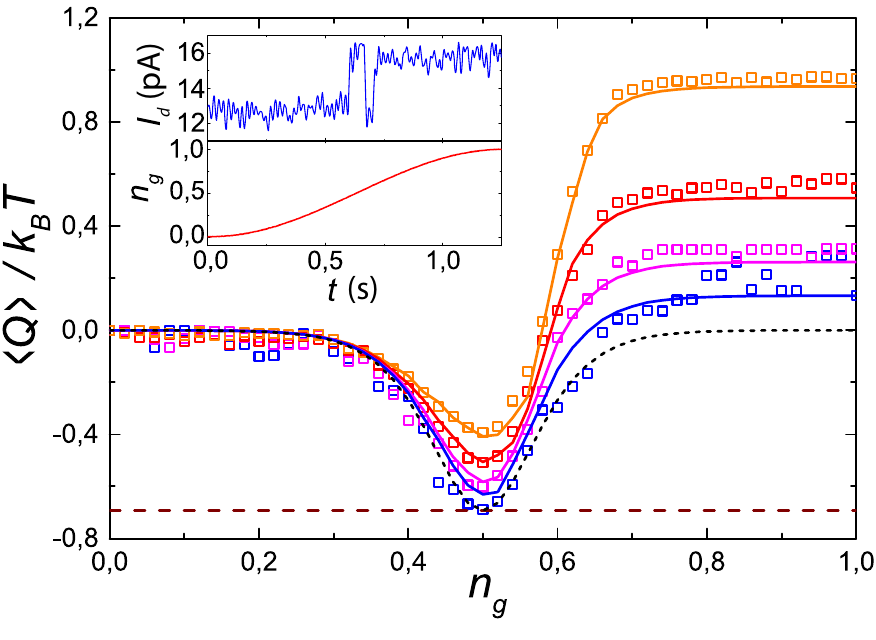}
\caption{
{Quasistatic drive.} The average total heat transferred to the reservoir in a ramp starting from $n_g = 0$ up to $n_g$ indicated on the $x$ axis. Symbols show the measured and solid lines the theoretical results. See supplementary material for details about the theoretical model (for all figures). Dashed curve gives the fully quasistatic limit of Eq.~~\ref{heat}, dashed straight line - the fundamental $-k_B T\ln 2$ limit. The maximum drive rates are $\dng = 0.22\Gamma_0$ for orange, $0.11\Gamma_0$ for red, $0.055\Gamma_0$ for magenta and $0.027\Gamma_0$ for blue, where $\Gamma_0 = 22$ Hz is the tunneling rate at degeneracy. The averages are taken over $N$ = 2105,~1764,~333, and $160$ repetitions, respectively. Inset: an example of realization of the measurement. }
\label{fig:ReversibleDissipation}
\end{figure}

Figure \ref{fig:ReversibleDissipation} shows the results of the measurements that illustrate such an extraction of heat from the reservoir. We drive our SEB starting from $n_g = 0$ towards $n_g = 1$ at various rates $\dot n_g$ while monitoring $n$ continuously to measure the total dissipated heat $Q$. We see that as the rate of the drive decreases, the average dissipated heat approaches the prediction of Eq.~\eqref{heat}: $\langle Q\rangle$ tends to $-k_B T \ln 2$ for $n_g = 0.5$. This process can also be viewed as the reversal of the Landauer erasure of one bit of information, in which the system is driven from the degeneracy with two equally occupied state to one certain configuration. Such an erasure produces at least $k_B T\ln 2$ of heat as demonstrated explicitly by recent experiments on a colloidal bead controlled with optical tweezers \cite{Berut12}. Since the drive in Fig.~\ref{fig:ReversibleDissipation} starts with $n_g = 0$, such that the SEB is in a definite state $n=0$ and thus initially $S_{ch}=0$, the lowest curve in this plot approaching Eq.~\eqref{heat} can be viewed as direct measurement of the equilibrium entropy $S_{ch}$ of the system of the two charge states $n=0,1$. When the quasistatic ramp to $n_g = 0.5$, as illustrated in Fig.~\ref{fig:ReversibleDissipation}, is complemented with an ideal measurement and immediate feedback that follows our Szilard engine protocol, the SEB operates as a reversible Maxwell demon, abstract models of which have been discussed theoretically recently \cite{Horowitz11,Mandal12,Barato13}.

\begin{figure}
\includegraphics[width=\columnwidth]{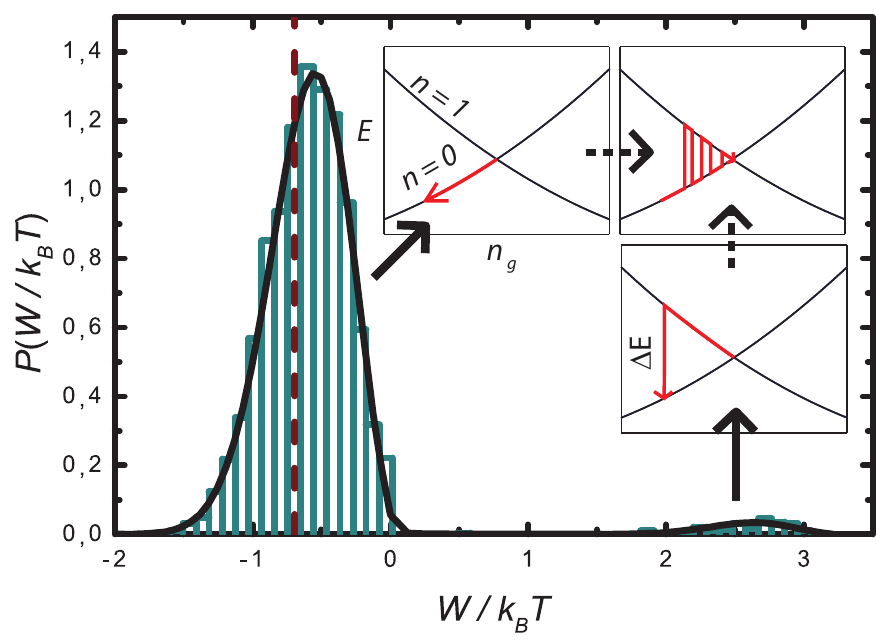}
\caption{{Distribution of work under feedback protocol.} The bars show the measured distribution, while the black line shows what is expected numerically. The inset panels display a sketch of the two processes corresponding to the two peaks in the distribution in the case $n$ was measured to be 0. Left panel shows a cycle with correctly performed feedback which contributes to the large peak at $W < 0$ around the ideal value $-k_B T \ln 2$ indicated by the dashed line. Cycles with
an error in the feedback (lower panel) send the box into the  large-energy state producing extra dissipation and contributing to the peak at $W > 0$. The overall work distribution shown here is obtained from $N = 2944$ cycles. The average extracted work for successful feedback response (the peak on the left hand side) is $\langle -W \rangle \approx 0.9 \times k_B T\ln(2)$, and the average of the full distribution is $\langle -W \rangle \approx 0.75 \times k_B T \ln(2)$. }
\label{fig:OptimalDistribution}
\end{figure}

Figure \ref{fig:OptimalDistribution} demonstrates the experimental performance of our Szilard engine, see supplementary material for details about the measurement protocol. Since the slow part of the cycle is not fully quasi-static, there are cycle-to-cycle fluctuations in $W$ (which is equal to $Q$ in each cycle), generating a distribution of $W$, which is obtained from a series of feedback cycle repetitions (as in Fig. \ref{fig:device}E). The cycles with correct gate-voltage feedback (left inset in Fig.~\ref{fig:OptimalDistribution}) trap the electron on the SEB island, on which it actually sits at degeneracy after the measurement. Then no electron tunneling occurs in the feedback process, and $W$ is close to the ideal limit $-k_B T\ln 2$. Such successful cycles produce the large peak at negative values of $W$ in Fig.~\ref{fig:OptimalDistribution}, around the ideal value that is indicated by the vertical dashed line. An error in the measurement or feedback drives the SEB to the excited charge state with excess energy $\Delta E = 2E_C |\Delta n_g|$, where $\Delta n_g$ is the total change in $n_g$ during the fast drive. Subsequent tunneling to the low-energy state (lower inset in Fig.~\ref{fig:OptimalDistribution}) dissipates energy $\Delta E \gg k_B T \ln 2$ extracted in the quasistatic part. Such cycles produce the small peak at positive values of $W$ in Fig.~\ref{fig:OptimalDistribution}.  For this measurement, we have chosen the optimized $|\Delta n_g| = 0.125$ in order to keep the contribution of the positive $W$ as small as possible, without significantly reducing the heat extracted from the thermal bath during the quasistatic drive. With this choice, we obtain an average extracted work per cycle of $\langle -W \rangle \approx 0.75 \times k_B T \ln 2$. For comparison, if no measurement were performed, only 50\% of the cycles would be successful, and one would do {\it positive} work $\langle W \rangle \approx 1.55 \times k_B T \ln2$ on the average.

To summarize, our experiment is a realization of a reversible Maxwell demon, similar to a Szilard engine, with a single electron box. We demonstrate quantitatively the extraction of $k_B T \ln 2$ of heat by creating a bit of information encoded in the position of the extra electron on one of the two islands of the box. Under a practical feedback cycle, our engine achieves a fidelity of about 75\%. The heat transfer measurements performed as a part of Maxwell demon demonstration provide also a direct measurement of the equilibrium entropy of a two-state system.

This work has been supported in part by the European Union Seventh Framework Programme INFERNOS (FP7/2007-2013) under grant agreement no. 308850, Academy of Finland (projects no. 139172, 250280, 272218), the National Doctoral Programme in Nanoscience, NGS-NANO (V.F.M.), and the NSF grant PHY-1314758 (D.V.A.). We acknowledge Micronova Nanofabrication Center and the Cryohall of Aalto University for providing the processing facilities and technical support.

\clearpage

\section{\Large Supplementary Material}

\section{Tunneling rates}

\begin{figure}[h!t]
\includegraphics[width=1\columnwidth]{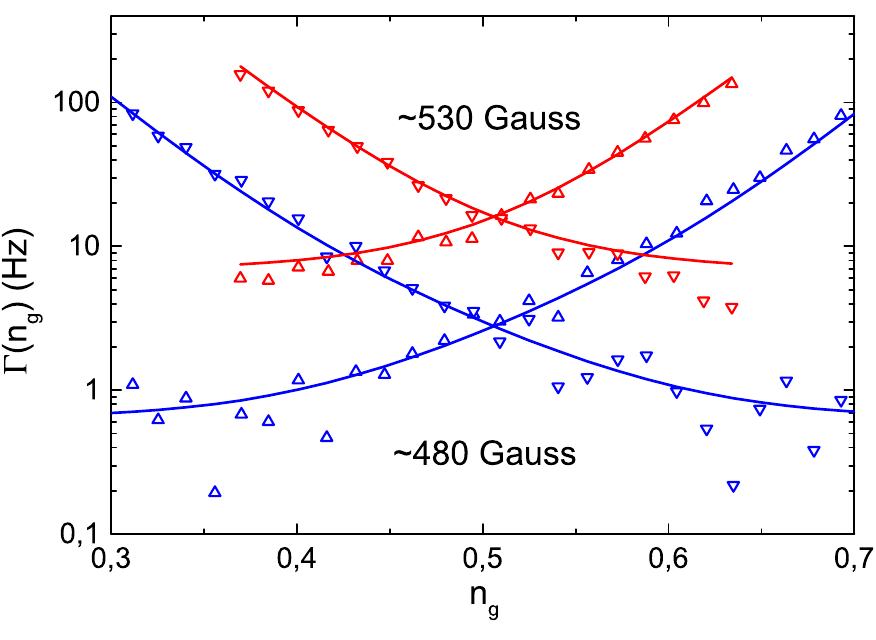}
\caption{Tunneling rates of the SEB at different magnetic fields at $T = 103$ mK. Triangles pointing up show the transition rates for $n: 0 \to 1$, and triangles pointing down show the transition rates for $n: 1 \to 0$. Solid lines show the rates given by Eq. \eqref{eq:TunnelingRate}. }
\label{fig:Rates}
\end{figure}

\begin{figure}[h!t]
\includegraphics[width=1\columnwidth]{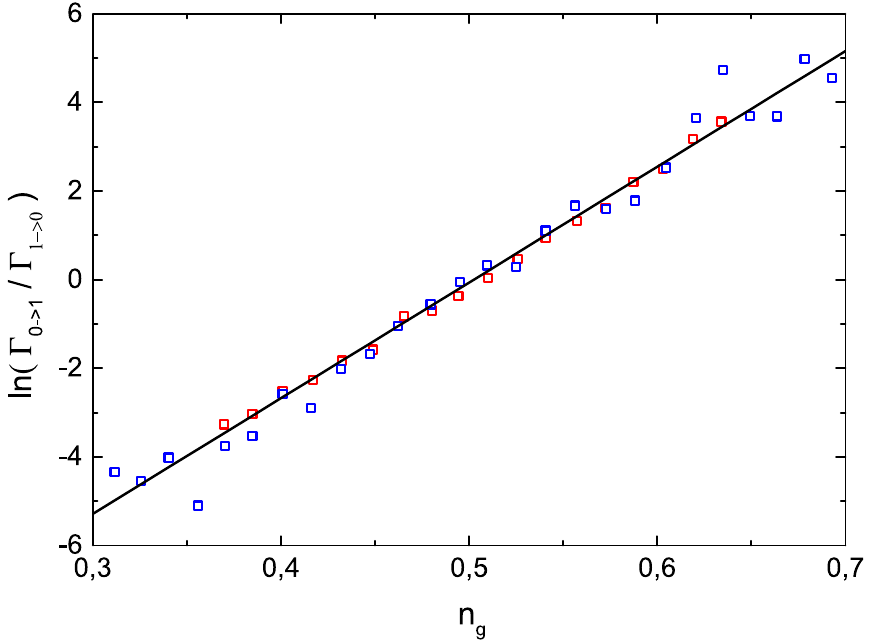}
\caption{$\ln(\Gamma_{0\to 1} / \Gamma_{1\to 0})$ of the rates shown in Fig. \ref{fig:Rates} as a function of $n_g$. 
The solid line shows a linear fit for $E_C / k_B T$ according to Eq. \eqref{eq:detailedbalance}.}
\label{fig:DetailedBalance}
\end{figure}

The tunneling rates of the SEB are given by
\be \Gamma = \frac{1}{e^2 R_t} \int dE N_S(E) f_S(E) f_N(-E -\Delta E), \label{eq:TunnelingRate} \ee
where $f_{S, N}(E)$ is the fermi distribution function of the $S$ or $N$ lead, $R_t$ is the tunneling resistance, and $N_S(E)$ is the normalized superconductor density of states. 
The change in electrostatic energy is $\Delta E = (2n_g-1)E_C/k_B T$ for the transition $n: 0\to 1$, and $\Delta E = -(2n_g-1)E_C/k_B T$ for $n: 1 \to 0$.
With $T_N = T_S = T$, the tunneling rates follow detailed balance 
\be \ln\left(\frac{\Gamma_{0\to 1}}{\Gamma_{1 \to 0}}\right) = \frac{\Delta E}{k_B T} = \frac{E_C}{k_B T} (2 n_g - 1), \label{eq:detailedbalance}\ee
where $\Gamma_{0\to1}$ and $\Gamma_{1 \to 0}$ are the rates of tunneling from states $n = 0$ to $1$ and vice versa.
The tunneling rate can be adjusted by an external magnetic field, effectively modifying the superconductor energy gap $\Delta$, as shown in Fig. \ref{fig:Rates}. 
We test the detailed balance condition by measuring the tunneling rates at different magnetic fields and temperatures, and check the slope of $\ln\left(\Gamma_{0\to 1} / \Gamma_{1\to 0}\right)$, as illustrated in Fig. \ref{fig:DetailedBalance}. Measuring the slope at different temperatures is consistent with Eq. \eqref{eq:detailedbalance}, validating detailed balance. The value of $E_C = 111$ $\mu$eV is independent of magnetic field (see Fig. \ref{fig:DetailedBalance}) and temperature.

\section{Measurement protocol}

The measurements are performed at the bath temperature of $T = 102 \pm 3$ mK, read by a RuO$_2$ thermometer. The tunneling rates are modified for each measurement by applying an external magnetic field to effectively modify the superconductor energy gap, see Supplementary material for details.
There is a $\sim 15$ ms delay between the measurement and the start of the fast ramp, including the delay due to numerical filtering. The drive reaches its extreme in 1 ms. For the measurement from which the distribution in Fig. 3 is obtained, $n_g$ is driven away by $0.125$ from degeneracy in the direction determined by the measurement outcome. This corresponds to an energy difference of $\Delta E \approx (3.4 \pm 0.2)k_B T$. The tunneling rate at degeneracy is approximately $\Gamma_0 = 1.3$ Hz, and the slow return back to degeneracy takes 10 s. %For testing the generalized Jarzynski equality, $n_g$ is driven away by $0.167$ from degeneracy, corresponding to an energy difference of $\Delta E \approx (4.4 \pm 0.2) k_B T$. The tunneling rate at degeneracy is approximately $\Gamma_0 = 2.7$ Hz, and the slow return is done in 8 s. The cut-off frequencies of the numerical filter for measurements are 100 Hz, 200 Hz, 300 Hz, 500 Hz, and 1 kHz, in the order of increasing error probability. We assume that $n$ is given by traces filtered with a 50 Hz cut-off frequency.

Fluctuating background charges influence $n_g$~\cite{Zimmerman}. Fortunately these changes are slow compared to the time scale of individual realizations of the experiment. 
Before driving the box, the gate is calibrated by applying a sinusoidal drive $n_g = n_{g, 0} + 0.5\cos(2 \pi f t)$, where $f$ is set to 5 Hz. Since the drive spans over a unit of $n_g$, a transition between $n = 0$ and $n = 1$ occurs during every half period of the drive. The drive offset $n_{g,0}$ is estimated based on the tunneling time instants, and is iteratively changed until $n_{g, 0}$ is close to 0.5. Simultaneously, the current levels $I_0$ and $I_1$ matching the states $n = 0$ and $n = 1$, respectively, are estimated from the histogram of the detector signal.

After the calibration of $n_g$, we move on to the actual feedback protocol. Initially, $n_g = 0.5$. We estimate $n$ by reading the detector signal, and by checking whether the latest data point is closer to $I_0$ or $I_1$. As soon as the state is estimated, $n_g$ is driven to $0.5 - \Delta n_g$ if $n = 0$ was measured, or to $0.5 + \Delta n_g$ if $n = 1$ was measured. Then $n_g$ is brought slowly back to degeneracy. The current levels $I_0$ and $I_1$ are re-evaluated from the histogram of the signal over the whole process, plus an additional 2 seconds spent at $n_g = 0.5$ in order to acquire sufficient statistics to estimate the current levels for both states, and to ensure that $\Ps$ follows thermal equilibrium distribution. 
We check the offset of $n_g$ according to the procedure described in the previous paragraph after a pre-set number (6...8) of repetitions. This is done to ensure that potential drifts of $n_g$ do not influence the result.
%In order to ensure the that $n_g$ is successively well calibrated, we check the calibration of $n_g$ according to the previous paragraph every pre-set number of repetitions, which for the experiments in the present manuscript was set to 6 for the measurement corresponding to Fig. 3, and 8 for the measurements corresponding to Fig. 4.

\begin{figure}[h!t]
\includegraphics[width=1\columnwidth]{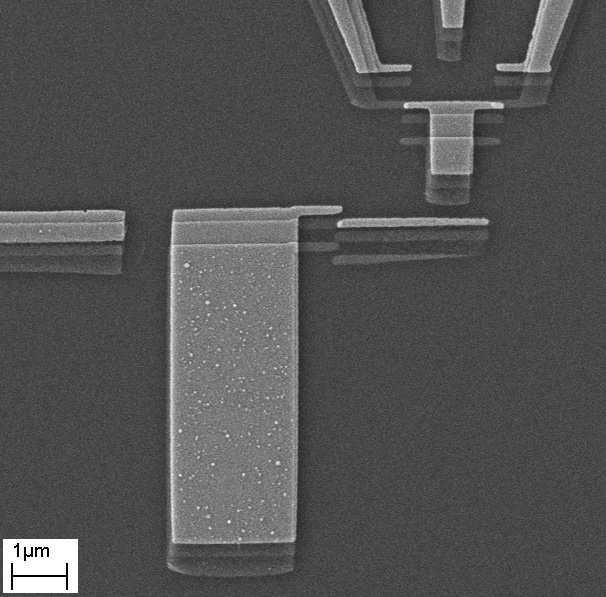}
\caption{An overall image of the single electron box (middle), the gate electrode (left), and the SET detector (top right).}
\label{fig:SampleImage}
\end{figure}

\section{Numerical simulation parameters}

The parameters used for the tunneling rates in Eq. \eqref{eq:TunnelingRate}, for the purpose of numerically estimating the distributions are listed below. 
$E_C = 111~\mu$eV and $R_t = 15$ M$\Omega$ are the same for all the simulations. $T$ is taken to be the temperature of the bath, and $\Delta$ is estimated to be approximately that measured for the detector at the given magnetic field strength.
Simulations related to Fig. 2 in the manuscript: $\Delta = 137$ $\mu$eV and $T = 103$~mK.
Simulations related to Fig. 3 in the manuscript: $\Delta = 157$ $\mu$eV and $T = 100$~mK.
The calculated average dissipated heat for Fig. 2 is obtained by solving the time evolution of $\Ps$ with the master equation. 
The numerical work distribution in Fig. 3 is obtained by calculating the time evolution of the characteristic function of $W$ using the techniques given in Ref. \cite{SairaThesis}

\section{Sample fabrication}

The sample fabrication and measurement techniques are similar to what is described in Refs. \cite{Saira, Koski},  First a layer of aluminium is deposited to form the first lead of the box, which is then oxidized with oxygen pressure of 90 mbar, while heating the sample stage in the evaporation chamber. The elevated temperature allows for higher junction resistance to slow down the tunneling rates in the actual experiment to a level measurable by a standard SET detector.% with a bandwith of ~100 Hz.
The chamber is allowed to cool down and the oxygen is then removed, after which the second layer of aluminium is deposited in a different angle to form the source and drain leads of the SET.
The sample is oxidized for the second time, with oxygen pressure of 2 mbar with oxidation time of 2 minutes to form the tunnel junctions of the SET. Last, a 30 nm layer of copper is deposited to form the island of the SET and the second island of the SEB. The aluminium island of the SEB is deliberately covered with a copper layer to improve its thermal relaxation, as shown in Fig. \ref{fig:SampleImage}.

\end{document}